\documentclass{jpp}

\usepackage{color}
\usepackage{graphicx}
\usepackage{natbib}

\ifCUPmtlplainloaded \else
  \checkfont{eurm10}
  \iffontfound
    \IfFileExists{upmath.sty}
      {\typeout{^^JFound AMS Euler Roman fonts on the system,
                   using the 'upmath' package.^^J}%
       \usepackage{upmath}}
      {\typeout{^^JFound AMS Euler Roman fonts on the system, but you
                   dont seem to have the}%
       \typeout{'upmath' package installed. JPP.cls can take advantage
                 of these fonts, if you use 'upmath' package.^^J}%
      }
  \else
  \fi
\fi

\ifCUPmtlplainloaded \else
  \checkfont{msam10}
  \iffontfound
    \IfFileExists{amssymb.sty}
      {\typeout{^^JFound AMS Symbol fonts on the system, using the
                'amssymb' package.^^J}%
       \usepackage{amssymb}%
         \let\leq=\leqslant
         
      }{}
  \fi
\fi

\ifCUPmtlplainloaded \else
  \IfFileExists{amsbsy.sty}
    {\typeout{^^JFound the 'amsbsy' package on the system, using it.^^J}%
     \usepackage{amsbsy}}
    {}
\fi

\newsavebox{\astrutbox}
\sbox{\astrutbox}{\rule[-5pt]{0pt}{20pt}}

\title[Kinetic inhibition of MHD-shocks in the vicinity of a parallel magnetic field]{Kinetic inhibition of MHD-shocks in the vicinity of a parallel magnetic field}

\author[A. Bret, A. Pe'er, L. Sironi, A. S\c{a}dowski and R. Narayan]%
{A\ls N\ls T\ls O\ls I\ls N\ls E\ns B\ls R\ls E\ls T$^{1,2,7}$, A\ls S\ls A\ls F\ns P\ls E'E\ls R$^3$, L\ls O\ls R\ls E\ls N\ls Z\ls O\ns S\ls I\ls R\ls O\ls N\ls I$^4$, A\ls L\ls E\ls K\ls S\ls A\ls N\ls D\ls E\ls R\ns S\ls \c{A}\ls D\ls O\ls W\ls S\ls K\ls I$^{5,6}$ and R\ls A\ls M\ls E\ls S\ls H\ns N\ls A\ls R\ls A\ls Y\ls A\ls N$^7$%
  \thanks{Email address for correspondence: antoineclaude.bret@uclm.es}
}

\affiliation{$^1$ETSI Industriales, Universidad de Castilla-La Mancha, 13071 Ciudad Real, Spain\\[\affilskip]
$^2$Instituto de Investigaciones Energ\'{e}ticas y Aplicaciones Industriales, Campus Universitario de Ciudad Real, 13071 Ciudad Real, Spain\\[\affilskip]
$^3$Physics Department, University College Cork, Cork, Ireland\\[\affilskip]
$^4$Department of Astronomy, Columbia University, New York, NY, 10027, USA\\[\affilskip]
$^5$MIT Kavli Institute for Astrophysics and Space Research, 77 Massachusetts Ave, Cambridge, MA 02139, USA\\[\affilskip]
$^6$Einstein Fellow\\[\affilskip]
$^7$Harvard-Smithsonian Center for Astrophysics, 60 Garden Street, MA 02138, USA
}

\date{?; revised ?; accepted ?. - To be entered by editorial office}
\begin{document}

\maketitle

\begin{abstract}
According to magnetohydrodynamics (MHD), the encounter of two
collisional magnetized plasmas at high velocity gives rise to shock
waves. Investigations conducted so far have found that the same
conclusion still holds in the case of collisionless plasmas.  For the case of a flow-aligned field, MHD stipulates that the field and the fluid are disconnected, so that the shock produced is independent of the field.
We present a violation of this MHD prediction when  considering
the encounter of two cold pair plasmas along a flow-aligned magnetic
field. As the guiding magnetic field grows, isotropization is
progressively suppressed, resulting in a strong influence of the field on the resulting structure. A micro-physics
analysis allows to understand the mechanisms at
work. Particle-in-cell simulations also support our conclusions and
show that the results are not restricted to a strictly parallel field.
\end{abstract}


\section{Introduction}
When a shock propagates into a neutral fluid, upstream particles slow
down at the shock front as a result of collisions with particles in
the slower-moving downstream gas. In fact, binary collisions are the
only possible microscopic mechanism for an upstream particle to slow
down. As a consequence, the shock front is a few mean-free-paths thick
\citep{Zeldovich}.

In-situ measurements of the earth's bow-shock within the solar wind
show that its front is far smaller than the mean-free-path of
the ions at the same location, which is comparable to an
astronomical unit \citep{PRLBow1, PRLBow2}. Such shocks, where the
mean-free-path is much larger than the front, have been dubbed
``collisionless shocks''. Instead of being sustained by binary
collisions, these shocks are mediated by collective plasma effects
acting on much shorter time and length scales than binary Coulomb
collisions \citep{Petschek1958,Sagdeev66}.

Collisionless shocks are believed to occur in a wide variety of
astrophysical settings: active galactic nuclei, pulsar wind nebulae,
planetary environments, supernova remnants, etc. The absence of
collisions allows particles to gain energy without sharing it
immediately with other particles. As a result, such shocks have been found
to be excellent particle accelerators and now count among the main
candidates for the production of high energy cosmic rays
\citep{Blandford1987,SironiReview2015, Marcowith2016}. They are also
believed to play a role in the generation of gamma-ray-bursts
\citep{Meszaros2014,Peer2015} and fast radio bursts
\citep{Lyubarsky2014,Falcke2014}.

Starting with the pioneering work of Sagdeev in the 1960's
\citep{Sagdeev66}, our knowledge of collisionless shocks has grown
tremendously, particularly in the past decade thanks to the advent of
large scale particle-in-cell (PIC) simulations \citep{Spitkovsky2005,
  Martins2009}. However, as recently as the 1990's, there were still
doubts about the very existence of collisionless shocks
\citep{Sagdeev_Kennel_1991}. While the earth bow shock measurements
have definitely eliminated these doubts, the micro-physics of
collisionless shock formation, and the mechanism of particle
acceleration, are still under investigation.

Given the omnipresence of collisionless shocks and their important
role in many phenomena, especially in astrophysics, the conditions for
such shocks to form are worthy of
investigation. A detailed understanding is all the more important
that electrostatic collisionless shocks\footnote{Before they collide,
  two plasmas display a Debye sheath at their border, with an
  associated potential jump \citep{gurnett2005}. At low energy of
  collision, the encounter is mediated by the interaction of these
  sheaths, and an electrostatic shock is formed. At higher energy, the
  interaction is rather mediated by the counter-streaming
  instabilities arising from the overlapping of the plasmas \citep{Stockem2013,BretJPP2015}. If the dominant instability is the Weibel one (see conditions in \cite{BretPoP2010}), then a ``Weibel shock''  is formed.}
have been observed in the laboratory \citep{Ahmed2013}, while the
production of Weibel mediated shocks such as the ones discussed here,
is expected within the next few years
\citep{Huntington2015,lobet2015,Park2016JPhC}. Note that the ``Weibel instability'' we refer to  is sometimes labelled ``filamentation instability'' of ``beam Weibel'' instability \citep{Silva2002,hill2005,DeutschPRE2005}. It is the instability of two counter-streaming flows with respect to perturbations with wave vectors normal to the flow.

When a collisionless shock forms from the encounter of two plasma
shells, the downstream plasma may be thermalized by collisionless
processes (see \cite{BretJPP2015} and references therein). As a
consequence, the equations of magnetohydrodynamics (MHD) can be
applied, so that both collisionless shocks and MHD shocks can in principle be
analysed using the same fluid approach\footnote{Once source of discrepancies are
  the accelerated particles which escape the Rankine-Hugoniot
  ``budget'' \citep{Stockem2012, Sironi2013, Caprioli2014,
    BretJPP2015}.}. For the case of a flow-aligned field, MHD prescribes that the fluid and the field are decoupled \citep{Majorana1987}, so that the very same shock should form, regardless of the field intensity.

Here, we present a specific example of departure from this expected MHD
behaviour. We consider the encounter of two collisionless cold pair
plasmas. A flow-aligned magnetic field is present, and the system is
relativistic. In section \ref{sec:mhd}, we explain the predictions of
MHD for this system. Then, in section \ref{sec:pic}, we describe a
series of simulations using the particle-in-cell (PIC)
technique. These simulations work at the microscopic level, and show a
departure from the MHD predictions beyond a critical magnetization. In
section \ref{sec:micro}, we present a micro-physics analysis of the
shock formation process  explaining the departure from
MHD.

 \begin{figure}
  \begin{center}
  \includegraphics[width=.45\textwidth]{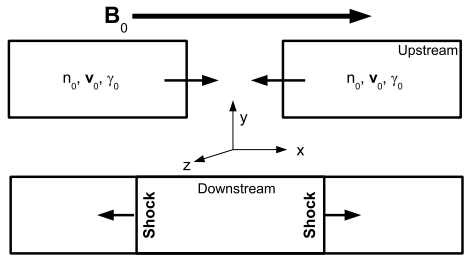}
  \end{center}
  \caption{Setup of the system considered. Two collisionless cold pair plasmas of density $n_0$ and initial Lorentz factor $\gamma_0$ collide over a flow-aligned magnetic field $\mathbf{B}_0$.}\label{fig:setup}
 \end{figure}

\section{System considered}
The system considered is shown schematically in
Fig.~\ref{fig:setup}. Two identical pair plasma shells of density $n_0$
head toward each other with initial velocity $\pm \textbf{v}_0$ and
Lorentz factor $\gamma_0 = (1-v_0^2/c^2)^{-1/2}$. The whole system is
embedded in an external field $\mathbf{B}_0 \parallel \mathbf{v}_0$
and aligned with the $x$ axis. We denote by ``upstream frame'' the
frame of reference of the right shell, and by ``downstream frame'' the
frame where the total momentum is 0. When a shock forms, these
frames become the upstream and downstream frames of the shock,
respectively. The strength of the magnetic field is measured by the
magnetization parameter,
\begin{equation}\label{eq:sigma}
\sigma = \frac{B_0^2/4\pi}{\gamma_0 n_0 m c^2},
\end{equation}
where all quantities are measured in the downstream frame.

\section{MHD predictions}\label{sec:mhd}
An MHD plasma sustains 3 kinds of modes: slow mode, Alfv\'{e}n mode,
and fast mode \citep{Kulsrud2005}. The phase velocities of these modes
satisfy the hierarchy $v_{\rm slow} < v_{\rm Alfven} < v_{\rm
  fast}$. Because of this hierarchy, the Alfv\'{e}n mode is sometimes
dubbed the ``intermediate mode'' \citep{Kulsrud2005}. In the cold
limit considered here, $v_{\rm slow}\rightarrow 0$ and $v_{\rm fast}
\rightarrow v_{\rm Alfven}$.

A ``fast shock'' has its front moving faster than the upstream fast
mode, while a ``slow shock'' only moves faster than the upstream slow
mode. For fast shocks, the shock front also propagates faster than the
downstream Alfv\'{e}n speed; in slow shocks, it propagates slower. An
intermediate regime exist, where the flow is super-Alfv\'{e}nic
upstream and sub-Alfv\'{e}nic downstream (crossing of the Alfv\'{e}nic
point, see eg \cite{Kirk1999}). However, such solutions of the MHD
jump equations do not survive when produced and are called
``extraneous'' \citep{Kulsrud2005}; they typically split into a pair
of ``fast'' and ``slow'' shocks.

For a flow-aligned field, the fluid motion decouples from the field
\citep{Majorana1987}. The shock formed is therefore the same,
regardless of the magnetization parameter $\sigma$. Nevertheless, its
front velocity can still be compared to the phase speeds of the three
modes. In the present cold limit, and for $\gamma_0\to\infty$, the shock
is expected to be ``fast'' for $\sigma < 2/3$, ``slow'' for $\sigma >
2$, and ``extraneous'' in between (see Appendix \ref{ap:MHD}). Figure \ref{fig:sigmagamma} shows these limits for a
range of $\gamma_0$ in the $(\sigma,\gamma_0)$ plane.

The MHD predictions for the present system  are therefore very
clear: the same shock should form regardless of the $\sigma$ parameter, simply because the fluid and the field are perfectly decoupled here. The MHD simulations run in Appendix \ref{ap:MHD} confirm this conclusion.

 \begin{figure}
  \begin{center}
  \includegraphics[width=.45\textwidth]{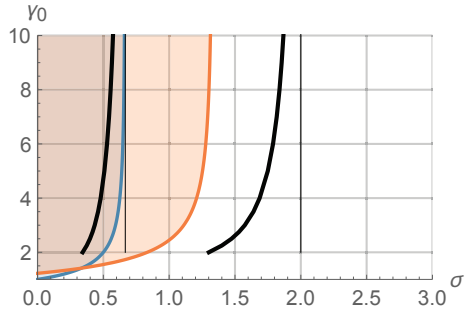}
  \end{center}
  \caption{The MHD thresholds for slow and fast shocks are represented
    by the thick black lines, with the thin vertical lines showing the
    large $\gamma_0$ limits, $\sigma=2/3$, 2. Extraneous shocks occur
    in between the two thick black lines. The Weibel instability
    governs systems located above and to the left of the orange
    curve. The Weibel filaments at saturation are able to stop the
    incoming flow, and initiate shock formation, only for systems to
    the left of the blue curve [Eq.~(\ref{eq:CritMicroB0}) with
      $\kappa=2/3$].}\label{fig:sigmagamma}
 \end{figure}

\section{PIC simulations}\label{sec:pic}
We now turn to PIC simulations to conduct a micro-physical, i.e.,
kinetic, analysis of the system under scrutiny. We use the 3D
electromagnetic PIC code TRISTAN-MP \citep{Spitkovsky2005}, which is a
parallel version of the publicly available code TRISTAN
\citep{Buneman1993} that has been optimized for studying relativistic
collisionless shocks
\citep{spitkovsky_08,spitkovsky_08b,sironi_spitkovsky_09,sironi_spitkovsky_11a,Sironi2013}. We
employ simulations in 2D computational domains, but all three
components of particle velocities and electromagnetic fields are
tracked (see more details in Appendix \ref{ap:PIC}).

\begin{table}
\begin{center}
\begin{tabular}{lcccccccc}
  $\sigma$   & 0.2 & 0.4 & 0.6 & 0.8 & 1   & 1.5 & 2   & 3 \\
  $\Omega_B$ & 2.0 & 2.8 & 3.5 & 4.0 & 4.5 & 5.5 &6.3  & 7.7
\end{tabular}
\end{center}
 \caption{Values of the parameter $\Omega_B=\sqrt{2\gamma_0\sigma}$ used in \cite{BretPoP2016} corresponding to the $\sigma$'s sampled here and for $\gamma_0=10$.}\label{tab:OmegaB}
\end{table}

We probe the regime $\gamma_0=10$ and $0 < \sigma < 3$. Note that the parameter space in \cite{BretPoP2016} is parameterized is terms of $\gamma_0$ and $\Omega_B$,  the later being related to the present $\sigma$ through $\Omega_B = \sqrt{2\gamma_0\sigma}$\footnote{The factor 2 comes from the fact that the plasma frequency in \cite{BretPoP2016} is the one of the electrons (or the positrons) alone, while the density $n_0$ in $\sigma$ is the total density of one pair beam.}. For better clarity, Table \ref{tab:OmegaB} gives the values of the $\Omega_B$'s of \cite{BretPoP2016} corresponding to the $\sigma$'s sampled here.

 \begin{figure}
  \begin{center}
   \includegraphics[width=.45\textwidth]{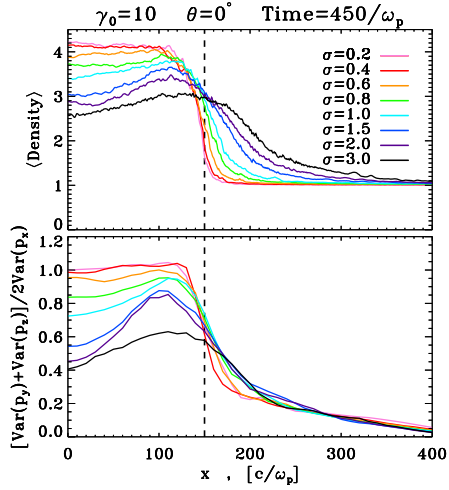}
  \end{center}
  \caption{Shock structure from a series of 2D PIC simulations with
    $\gamma_0=10$, $0.2 \leq \sigma \leq 3$, at $\omega_p t
    =450$. We plot the $y$-averaged density profile (top panel) and a
    measure of the plasma anisotropy (bottom panel), as defined in
    Eq.~(\ref{eq:var}). The vertical dashed line indicates the
    position of the front, assuming that it propagates
    at $c/3$. The angle between the field and the flow is
    $\theta=0$.}\label{fig:gamma10}
 \end{figure}

Figure \ref{fig:gamma10} shows the $y$-integrated density profile of
the system for $\gamma_0=10$ (top panel), at a relatively early time,
$\omega_p t =450$, where $\omega_p^2=4\pi n_0 q^2/\gamma_0
m$. The magnetization parameter varies from 0.2 to 3, as indicated in
the legend. In the bottom panel, we quantify the isotropization of the
particle distribution function by plotting the ratio $\varphi$ between the
momentum dispersion along the transverse directions ($y$ and $z$) as
compared to the longitudinal direction $x$, namely,
\begin{equation}\label{eq:var}
  \varphi = \frac{\mathrm{Var}(p_y) + \mathrm{Var}(p_z)}{2\mathrm{Var}(p_x)}.
\end{equation}
We notice that, for $\sigma\lesssim 0.4$, the shock structure is
independent of the magnetization, in line with the MHD predictions. In the downstream (left of the
vertical dashed line), the density approaches the value predicted by
the MHD jump conditions ($\sim 4.2$ for $\gamma_0=10$, and $\sim 4$ in
the limit $\gamma_0\gg 1$). Correspondingly, the shock speed
approaches the value $\sim c/3$ predicted by MHD (indicated by the
dashed line). The bottom panel in Fig.~\ref{fig:gamma10} shows that
for low magnetizations the downstream plasma is nearly isotropic. However, for
higher magnetizations ($\sigma\gtrsim 0.6$), the downstream density is
lower than the value predicted by MHD. Consequently, the shock speed
is faster than the MHD prediction $\sim c/3$.

Noteworthily, the width of the density jump increases notably with $\sigma$. For small values, the shock front is $\sim 70 c/\omega_p$ thick. But for $\sigma=3$, the transition region between the ``upstream'' and the ``downstream'' is $\sim 300 c/\omega_p$.

Why do the results for $\sigma\gtrsim 0.6$ deviate from MHD? One might
think then, that because the PIC simulations are limited to early times,
the shock has not formed yet. How much time should the
formation of a shock take? For the present system, the growth-rate
$\delta_W$ of the Weibel instability is given by
\citep{StockemApJ2006,BretPoP2016},
\begin{equation}\label{eq:grW}
\delta_W = \omega_p \sqrt{2\beta_0^2-\sigma},
\end{equation}
where $\beta_0=v_0/c$. The shock formation time typically amounts to a few tens of
$e$-folding times \citep{BretPoP2013,BretPoP2014}. With the parameters
used here, $20\delta_W^{-1}$ is at most $28\omega_p^{-1}$ for
$\sigma=1.5$ ($\delta_W$ vanishes for $\sigma >
2\beta_0^2$). Therefore, the time $t=450\omega_p^{-1}$ to which
the simulations in Fig.~\ref{fig:gamma10} have been run, exceeds by a
factor of 15 the slowest expected shock formation time.

  \begin{figure}
  \begin{center}
   \includegraphics[width=.45\textwidth]{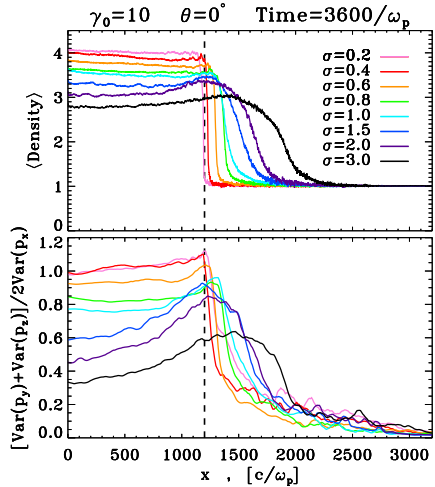}
  \end{center}
  \caption{Same as in Fig.~\ref{fig:gamma10}, but at a later time:
    $\omega_p t =3600$.}\label{fig:gamma10_long}
 \end{figure}

To verify the above argument, we have evolved the simulations to much
longer times: $\omega_pt=3600$ (Fig.~\ref{fig:gamma10_long}). We again
find that the density profile strongly varies with $\sigma$, contrary to the MHD prescriptions. For magnetizations $\sigma\gtrsim 0.6$, the system settles
in a quasi-stationary state which does not satisfy the usual MHD jump
conditions. Ultimately, the fact that the density jump and the shock
speed do not agree with the MHD jump conditions is related to the lack
of isotropy in the downstream plasma. As shown in the bottom panel of Figs. \ref{fig:gamma10} \& \ref{fig:gamma10_long} , for $\sigma\gtrsim 0.6$, the downstream particle distribution is hotter
along the longitudinal direction than in the transverse
directions\footnote{Note that, since the particle momenta are measured in the
downstream frame of the simulations, we do not expect $\varphi=1$ in
the upstream medium (but rather $\varphi\propto 1/\gamma_0$), despite
the fact that the upstream plasma is isotropic in its own rest
frame.}. For large $\sigma$, we find downstream $\varphi<1$ (left of the
vertical dashed lines), which means that the flow is not isotropized even at late times.

The width of the density jump is again worth emphasizing. For small values of $\sigma$, the shock front on Fig. \ref{fig:gamma10_long} is still $\sim 70 c/\omega_p$ thick. But for $\sigma=3$, the transition region is now $\sim 2000 c/\omega_p$.

The micro-physical analysis discussed next in
section \ref{sec:micro} predicts that the departure from the MHD behavior we just observed, is $\gamma_0$-independent at large $\gamma_0$. This prediction has  been successfully
tested in Appendix \ref{ap:PIC} by running a series of PIC simulations with $\gamma_0=30$.
We also confirm in Appendix \ref{ap:PIC} that these results are not restricted to a
perfectly flow-aligned field ($\theta=0$) but survive even for a
misaligned field.

\section{Micro-physics of the shock formation}\label{sec:micro}
From the discussion so far, it appears that the observed departure of
the system under consideration from the predictions of MHD, boils down
to the non-isotropization of the downstream particle
distribution function, even at late times. The following kinetic
analysis of the shock formation process allows us to understand why
isotropization fails.

Weibel shocks are mediated by purely collective phenomena. When the
two plasma shells start interpenetrating, the overlapping region turns
unstable to counter-streaming instabilities. Many linear instabilities
compete \citep{BretPoP2010}, but the Weibel (filamentation)
instability, with a $\mathbf{k}$ normal to the flow, grows faster than
all others, provided \citep{BretPoP2016},
\begin{equation}\label{eq:Wwins}
 \gamma_0 > \sqrt{\frac{2}{4/3-\sigma }}.
\end{equation}
The line corresponding to this limit is shown in
Fig.~\ref{fig:sigmagamma} by the orange curve. The Weibel instability
dominates the unstable spectrum of the system for all points of the
$(\sigma,\gamma_0)$ plane above and to the left this line.

The micro-physics of shock formation depends on the ability of the
Weibel instability to form magnetic filaments capable of blocking the
plasma that keeps entering the overlapping region. In the case of
un-magnetized pair plasmas, for example, this condition is already met
at saturation of the Weibel instability
\citep{BretPoP2013,BretPoP2014}. As a result, the density quickly
builds up in the overlapping region, and a shock forms. Distribution
functions are quickly isotropized in the overlapping region, and MHD
considerations apply.

The magnetic filaments generated by the Weibel instability are of the
form,
\begin{equation}\label{eq:Wfield}
\mathbf{B}_f = B_f \sin(k\,y)  ~ \mathbf{e}_z,
\end{equation}
where $k$ is the fastest growing wave-number. When there is no
external magnetic field, an analysis of the motion of a particle of
mass $m$ and charge $q$ in such filaments \citep{BretPoP2015} shows that it is stopped
inside if
\begin{equation}\label{eq:CritMicro}
k^{-1} > \frac{v_0}{\omega_{B_f}},~~\mathrm{with}~~\omega_{B_f} = \frac{q B_f}{\gamma_0 m c},
\end{equation}
where $\mathbf{v}_0$ is the initial, flow-aligned velocity of the
particle and $\gamma_0$ is its Lorentz factor. Although the model from
which this conclusion is derived is highly simplified, the condition
is consistent with the results of PIC simulations \citep{BretPoP2014}.

How is Eq.~(\ref{eq:CritMicro}) modified in the presence of a
flow-aligned magnetic field? One would expect a guiding field to
suppress the transverse scattering of particles and to thereby help
particles go through the filaments without stopping. Indeed, analysis
shows that regardless of their initial velocity or initial position
along the $y$ axis, all particles stream through the filaments
whenever \citep{BretJPP2016}
\begin{equation}\label{eq:crit0}
B_0 > \frac{1}{2}B_f.
\end{equation}
Since $B_f$ arises from the growth of the Weibel instability, its
magnitude can be quantified
\citep{StockemApJ2006,BretPoP2016}. Therefore, the above criterion can
eventually be expressed in terms of $\sigma$ and $\beta_0^2 =
1-1/\gamma_0^2$, giving (see details in Appendix \ref{appen1}),
\begin{equation}\label{eq:CritMicroB0}
  \sigma > \kappa\beta_0^2,
\end{equation}
where $\kappa=2/3$ if equipartition is assumed at saturation of the
Weibel instability. The boundary
corresponding to the criterion (\ref{eq:CritMicroB0}) with $\kappa =
2/3$ is shown in Figure \ref{fig:sigmagamma} by the blue curve.

The region between the bounds corresponding to Eqs. (\ref{eq:Wwins})
and (\ref{eq:CritMicroB0}), i.e., the region between the orange and
blue lines in Fig. \ref{fig:sigmagamma}, corresponds to a range of
parameters where the Weibel instability governs the linear phase of
the overlapping region, but the filaments at saturation are not strong
enough to stop the flow. The expected consequence, as indeed observed
in our PIC simulations, is that the flows are not trapped in the
overlapping region, but keep streaming through. Isotropization is not
achieved, and MHD does not apply.

The reader my have noticed that for $\sigma=2$ and 3, the Weibel instability does \emph{not} govern the linear phase of the initial interaction between to two shells. How is it then that the system still fails to
follow MHD? We conjecture that the analysis described above,
where we were able to quantify all the steps because the Weibel
instability is well understood, must be a particular case of the
following more general argument. Instead of the Weibel filaments
described by Eq.~(\ref{eq:Wfield}), consider a turbulent
electromagnetic perturbation $\sum_\mathbf{k} \mathbf{E}_\mathbf{k} +
\mathbf{B}_\mathbf{k}$ (with $<\mathbf{E}_\mathbf{k}> =
<\mathbf{B}_\mathbf{k}> =\mathbf{0}$) that is present in the
overlapping region and that can potentially isotropize the incoming
flow. Consider also a superimposed, flow-aligned field
$\mathbf{B}_0$. In the limit $B_0=0$, the incoming flow is
isotropized, and usual MHD applies. In the opposite limit $B_0\to\infty$,
the incoming flow is strongly guided by the mean field, and will
ignore the weaker underlying turbulence. Hence, MHD prescriptions are violated. When
does the switch from one regime to the other happen? We conjecture
that particles will tend to follow the mean field instead of being
randomized whenever the energy density $B_0^2/8\pi$ of the mean field
exceeds a fraction of order unity of the turbulent energy
$\mathcal{E}_T$. Now, if the turbulence is caused by an instability of
the counter-streaming flows, its energy will be a fraction of the flow
energy density, i.e., $\mathcal{E}_T \lesssim \gamma_0 n_0 m c^2$. As
a consequence, the system will depart from MHD beyond a critical value of
$(B_0^2/8\pi)/\gamma_0 n_0 m c^2 = \sigma/2$. We thus conclude that,
regardless of which instability is initially triggered in the
overlapping region, the MHD behaviour is inhibited for values of
$\sigma$ greater than about unity. This is indeed what is observed in our
PIC simulations.

\section{Conclusions}
In summary, we have found a departure from MHD behaviour when two
collisionless pair plasma shells with a flow-aligned magnetic field
collide. While MHD stipulates that the very same shock should form regardless of the $\sigma$ parameter, the micro-physics analysis of the shock formation allows to understand why the standard shock formation scenario can be jeopardized beyond a critical magnetization.

PIC simulations have confirmed the theoretical analysis.  The results are similar when considering an angle $\theta=5^\circ$ between the field and the flow (see Appendix \ref{ap:PIC}). This shows that the observed MHD departure is not a ``Dirac delta'' effect, strictly restricted to $\theta=0$.

What about an electron/proton plasma? It is difficult at this stage to draw definite conclusions about that case. When protons are accounted for instead of positrons, the asymmetric role of electrons and protons results in an upstream current which, in the presence of a flow-aligned magnetic field, is likely to trigger the Bell instability \citep{Bell2004}. This instability is not triggered here because of the symmetric role of electrons and positrons. But if excited, the upstream Bell turbulence, when transported downstream, could help isotropizing the flow. Yet, in spite of some differences with pair plasmas \citep{Stockem2015ApJ}, shock formation in electron/proton plasmas eventually still boils down to the capability of an instability generated turbulence to stop the flow. If the conjecture enounced at the end of Section \ref{sec:micro} turns out to be valid, we could recover a $\sigma$ threshold for the validity of MHD in electron/proton plasmas as well, since the energy of the downstream turbulence should remain a fraction of the upstream kinetic energy. Further studies will be necessary to sort out this important issue.

Would it be possible to modify MHD so that it keeps fitting the kinetic results for $\sigma \gtrsim 0.6$? A tentative pathway, beyond the scope of this work, would  be to include the downstream anisotropy within the MHD analysis. Indeed, bottom-Figs. \ref{fig:gamma10} \& \ref{fig:gamma10_long} clearly show that the downstream is not isotropized because of the magnetic field. One could therefore try to quantify this anisotropy in terms of the field, before inserting the corresponding temperature anisotropy in the Rankine-Hugoniot jump conditions analysis \citep{Karimabadi95,Vogl2001,Gerbig2011}.

Future work will also explore in detail the angular dependence of our results, together with the expected consequences for astrophysics.

\section{Acknowledgements}
AB acknowledges grants ENE2013-45661-C2-1-P, PEII-2014-008-P and
ANR-14-CE33-0019 MACH. AP acknowledges support by the European Union
Seventh Framework Program (FP7/2007-2013) under grant agreement
\#618499, and support from NASA under grant \#NNX12AO83G. RN's
research was supported in part by NASA grant TCAN NNX14AB47G. OS
acknowledges support by NASA through Einstein Post-doctoral Fellowship
number PF4-150126 awarded by the Chandra X-ray Center, operated by the
Smithsonian Astrophysical Observatory for NASA under contract
NAS8-03060. Thanks to Smadar Naoz and Victor Malka for valuable
inputs.

\appendix

\section{MHD predictions}\label{ap:MHD}
\subsection{MHD criterion for slow shocks}
A key quantity is the speed of the shock front relative to the phase
velocity of the fast mode. Consider first the upstream fast mode. Its
phase velocity $v_{fast}$, which is also the phase velocity of the
Alfv\'{e}n mode, is given in the upstream frame by \citep{Kirk1999,
  Keppens2008}\footnote{The following expression allows for the relativistic corrections needed when the field energy density $B_0^2/8\pi$ approaches the matter energy density $n_0 m c^2$. In such circumstances, the field participates in the inertia of the medium and modifies the speed of the waves.},
\begin{equation}\label{eq:fms}
  v_{\rm fast}^2 = c^2\frac{\sigma'}{1+\sigma'},~~\mathrm{with}~~\sigma' = \frac{B_0^2}{4\pi n_0' m c^2},
\end{equation}
where the prime stands for quantities measured in the upstream
frame. Note that the field has no prime because it is aligned with the
axis of the Lorentz transformations.

Regarding the velocity of the shock front, it is in 2D and for large
$\gamma_0$, $\beta_s \equiv v_s/c =1/3$ in the downstream frame
\citep{Kirk1999}. In this frame, the upstream propagates at $\mathbf{v}_0$
(Fig.~\ref{fig:setup}). Therefore, the speed $v_s'$ of the shock front
in the upstream frame is
\begin{equation}\label{eq:vs}
  \frac{v_s'}{c} \equiv \beta_s' = \frac{\beta_s + \beta_0}{1 + \beta_s\beta_0} \sim 1-\frac{1}{4\gamma_0^2}~~\mathrm{if}~~\gamma_0 \gg 1.
\end{equation}
The condition $v_{\rm fast} > v_s'$ reads therefore ($\gamma_0 \gg
1$),
\begin{eqnarray}\label{eq:crit1}
  \frac{\sigma'}{1+\sigma'} &>& \left( 1-\frac{1}{4\gamma_0^2}\right)^2 \nonumber \\
  \Rightarrow  \sigma' &>& 2\gamma_0^2.
\end{eqnarray}

We need now to express the left-hand-side of this equation in terms of the $\sigma$ parameter Eq.~(\ref{eq:sigma}).
Since the field $\mathbf{B}_0$ is aligned with the axis of the Lorenz transformation, it does not change. Regarding the density, there is a  relativistic bunching between $n_0'$, the upstream density in the upstream frame, and $n_0$, the upstream density in the downstream frame, with
\begin{equation}\label{eq:nprime}
n_0 = \gamma_0 n_0'.
\end{equation}
Combining with Eqs. (\ref{eq:sigma},\ref{eq:crit1}), we finally obtain,
\begin{equation}\label{eq:crit}
  \sigma > 2.
\end{equation}
Therefore, MHD shocks with $\sigma > 2$ have to be of the slow type. This limit is pictured on Fig.~\ref{fig:sigmagamma} in the $(\sigma,\gamma_0)$ phase space, together with a refined calculation accounting for the $\gamma_0$-dependence of the threshold.

Our calculation has been conducted in 2D in order to compare with the  PIC simulations. In 3D, one has $\beta_s=1/4$ and Eq.~(\ref{eq:crit}) reads $\sigma > 5/3$ instead. We now turn to the same analysis, but for the downstream.

\subsection{MHD criterion for extraneous shocks}\label{sec:extra}
Measured in the downstream, the downstream Alfv\'{e}n and fast mode velocities, read
\begin{equation}\label{eq:fms2}
  \beta_{\rm fast}^2 = \frac{\sigma_w}{1+\sigma_w},
\end{equation}
with,
\begin{equation}
\sigma_w = \frac{B_0^2}{4\pi w},~~\mathrm{and}~~w = n m c^2 + \frac{\hat{\gamma}}{\hat{\gamma}-1}nk_BT,
\end{equation}
where $\hat{\gamma}=4/3$ is the adiabatic index, $w$ the enthalpy density, and $n$ the downstream density in its own frame. Writing \citep{Service1986ApJ},
\begin{equation}
n k_BT=\frac{4}{3}\gamma_0^2 n_0 m c^2,
\end{equation}
and neglecting $n m c^2$ in the enthalpy, we obtain
\begin{equation}
  \sigma_w = \frac{3}{16} \frac{B_0^2/4\pi}{\gamma_0^2 n_0 m c^2}.
\end{equation}
Since in the downstream frame, the shock front propagates at $c/3$, we need to compare $\beta_{\rm fast}^2$ with $1/9$. The equation $\beta_{\rm fast}^2=1/9$ gives,
\begin{equation}\label{eq:critX}
  \sigma_w = \frac{1}{8}~~\Rightarrow ~~ \frac{B_0^2/4\pi}{\gamma_0^2 n_0 m c^2}=\sigma = \frac{2}{3}.
\end{equation}
The corresponding $\sigma$-limit is pictured on the same Fig.~\ref{fig:sigmagamma}. Table \ref{tab:sigma} summarizes the MHD conditions for super-Alfv\'{e}nic upstream and downstream, Eqs.~(\ref{eq:crit},\ref{eq:critX}). As it appears, the interval $2/3<\sigma<2$ defines a range of extraneous solutions. For $\sigma < 2/3$, MHD gives a fast shock and for $\sigma>2$, it gives a slow shock.

\begin{table}
\begin{center}
\begin{tabular}{lccc}
  $\sigma$   & $<2/3$ & $<2$ & $>2$ \\
    \hline
  Upstream   & Super-A & Super-A & Sub-A \\
  Downstream & Super-A & Sub-A & Sub-A \\
  Shock type  & Fast    & Extraneous & Slow \\
\hline
\end{tabular}
\end{center}
 \caption{Summary of the MHD conditions for super-Alfv\'{e}nic upstream and downstream for $\gamma_0 \gg 1$, Eq.~(\ref{eq:crit}, \ref{eq:critX}). The interval $2/3<\sigma<2$ pertains to extraneous solutions.}\label{tab:sigma}
\end{table}

  \begin{figure}
  \begin{center}
   \includegraphics[width=.45\textwidth]{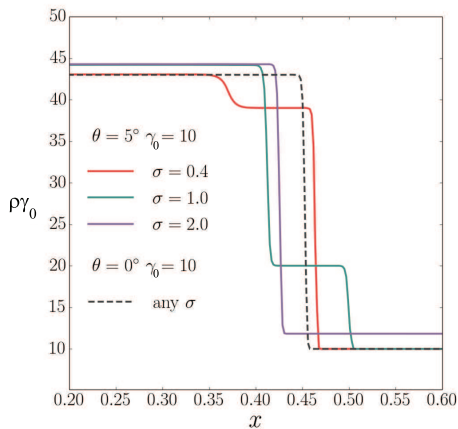}
  \end{center}
  \caption{MHD simulations of the collision of two cold pair plasmas with $\gamma_0=10$. The extraneous regime requires a tilted field in order to appear.}\label{fig:mhd}
 \end{figure}

\subsection{MHD simulations}\label{sec:mhdSim}
We performed 1D MHD simulations of colliding magnetized flows in order to confirm the previous analysis. We used the general relativistic magnetohydrodynamical code KORAL \citep{Sado2014}. The domain was filled initially with gas of uniform comoving frame density, $\rho=1$, and velocities set up so that the gas collides at $x=0$ and on both sides of the contact surface the velocities correspond to the given Lorentz factor $\gamma_0$. The gas magnetization is described by the $\sigma$ parameter and tilt angle $\theta$ (zero tilt angle corresponds to field perfectly parallel to the flow).
We ran a few simulations for the flow-aligned case ($\theta=0$) as well as for a $5^\circ$ tilt. The results are shown in Fig.~\ref{fig:mhd}. For the flow-aligned field, the exact decoupling of the fluid motion from the field \citep{Majorana1987} results in a shock independent from $\sigma$. For the $5^\circ$ tilt, the field couples to the fluid and the extraneous regime is unravelled as the shock splits into 2 sub-shocks \citep{Kulsrud2005} for intermediate $\sigma$ parameters.

\section{PIC simulations}\label{ap:PIC}
\subsection{Simulation Setup}

The shock is set up by reflecting a cold ``upstream''  flow from a conducting wall located at $x = 0$. The interaction between the incoming beam (that propagates along $-\mathbf{e}_x$) and the reflected beam triggers the formation of a shock, which moves away from the wall along $+\mathbf{e}_x$. This setup is equivalent to the head-on collision of two identical plasma shells  (Fig.~\ref{fig:setup}), which would form a forward and reverse shock and a contact discontinuity. Here, we follow only one of these shocks, and replace the contact discontinuity with the conducting wall.
The simulation is performed in the ``wall'' frame, where the ``downstream'' plasma behind the shock is at rest.

We use a rectangular simulation box in the $xy$ plane,
with periodic boundary conditions in the $y$ direction. The incoming plasma is injected through a ``moving injector,'' which recedes from the wall along $+\mathbf{e}_x$ at the speed of light. The simulation box is expanded in the $x$ direction as the injector approaches the right boundary of the computational domain. This permits us to save memory and computing time, while following the evolution of all the upstream regions that are causally connected with the shock.

In 2D, each computational cell is initialized with 16
electrons and 16 positrons. The relativistic electron skin depth for the incoming
plasma ($c/\omega_p$, with $\omega_p^2=4\pi n_0 q^2/\gamma_0 m$) is resolved with 10 computational
cells and the simulation time-step is $\Delta t = 0.045 ~\omega_p^{-1}$. The computational domain is typically $\sim 102 c/\omega_p$ wide (corresponding to 1024 cells). The simulations extend typically up to $ \omega_p t \sim 3600$, corresponding to a box length of $\sim 3600 c/\omega_p$, or $\sim 36000$ cells.

The  incoming stream is injected along $-\mathbf{e}_x$ with bulk Lorentz factor $\gamma_0$. The incoming plasma is cold, with thermal spread $k_BT/m c^2\ll1$. We vary the upstream Lorentz factor between $\gamma_0=10$ and 30, but we find that our results are nearly insensitive to the choice of the Lorentz factor, as long as the flow is ultra-relativistic. The upstream flow is seeded with a uniform magnetic field $\mathbf{B}_0$.

  \begin{figure}
  \begin{center}
   \includegraphics[width=.45\textwidth]{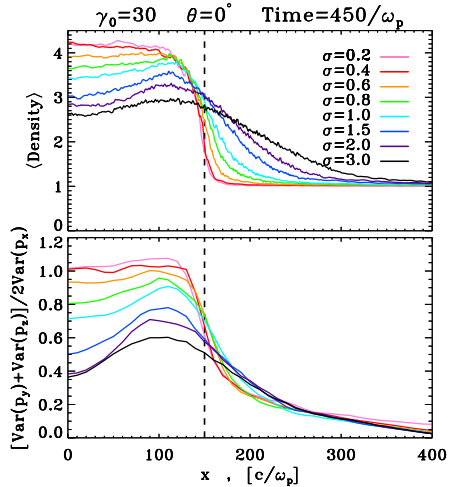}
  \end{center}
  \caption{Same as Fig.~\ref{fig:gamma10}, but for $\gamma_0=30$.}\label{fig:gamma30}
 \end{figure}

Figure \ref{fig:gamma30} shows the result of a series of simulations similar to those displayed on Fig.~\ref{fig:gamma10}, but for $\gamma_0=30$. The weak $\gamma_0$-dependence, expected from the micro-physics analysis of Sec. \ref{sec:micro}, is confirmed.

\subsection{Oblique Shocks}
In most of our studies, the upstream field is aligned with the flow (i.e., we study a ``parallel shock''), but we also consider quasi-parallel shocks in which the upstream field makes an angle $\theta=5^\circ$ with the flow direction of propagation\footnote{In 2D simulations, we initialize the magnetic field out of the plane of the simulations. Yet, we have verified that our results do not change for an in-plane field with the same obliquity $\theta$. In this case, we also initialize a motional electric field $\mathbf{E}_0=-\mbox{\boldmath{$\beta$}}_0\times\mathbf{B}_0$ in the upstream medium, where  $\mbox{\boldmath{$\beta$}}_0=-\beta_0\,\mathbf{e}_x$ is the three-velocity of the injected plasma.}.

\begin{figure}
  \begin{center}
   \includegraphics[width=.45\textwidth]{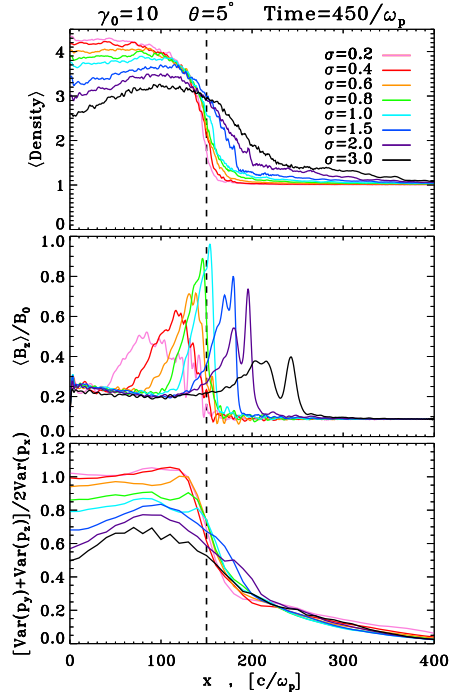}
  \end{center}
  \caption{Shock structure from a series of 2D PIC simulations with $\gamma_0=10$, $0.2 \leq \sigma \leq 3$,  at $\omega_p t =450$. We plot the $y$-averaged density profile (top panel), the $y$-averaged profile of the transverse field $B_z$ (middle panel), and a measure of the plasma anisotropy (bottom panel), as defined in the text. The vertical dashed line indicates the position of the shock front, assuming that it propagates at $c/3$. The angle between the flow and the field is $\theta=5^\circ$.}\label{fig:theta5}
 \end{figure}

In Fig.~\ref{fig:theta5}, we present the results of a suite of 2D simulations of nearly-parallel shocks, with obliquity $\theta=5^\circ$. In addition to the $y$-averaged profile of the particle number density (top panel) and the measure of particle anisotropy (bottom panel), we present the $y$-averaged profiles of the transverse magnetic field $B_z$, in units of the upstream field $B_0$. For such small field obliquity, most of the conclusions presented above for the case of parallel shocks still hold. In particular, for $\sigma\gtrsim 0.6$ the downstream flow is no longer isotropic. As a result, the density jump is lower than the MHD predictions.

\section{Value of the parameter $\kappa$ in Eq.~(\ref{eq:CritMicroB0})}\label{appen1}
The criterion (\ref{eq:CritMicroB0}) has been obtained in \cite{BretJPP2016} assuming equipartition at saturation of the Weibel instability. We here elaborate on the parameter $\kappa$ beyond what is done in \cite{BretJPP2016}.

For the present system, the growth-rate $\delta_W$ of the Weibel instability is given by Eq.~(\ref{eq:grW}). The instability grows the field given by Eq.~(\ref{eq:Wfield}) until \citep{davidsonPIC1972},
\begin{equation}\label{eq:equi}
  \frac{q \sqrt{<\mathbf{B}_f^2>}}{\gamma_0 m c}=\frac{q B_f/\sqrt{2}}{\gamma_0 m c} = \eta\delta_W ,
\end{equation}
where $\eta = 1$ means equipartition, as evidenced by Eqs. (\ref{eq:equiTot},\ref{eq:equiW}) below\footnote{The $\sqrt{2}$ factor in Eq. (\ref{eq:equi}) is not accounted for in \cite{BretJPP2016}}. The magnetic
filaments become unable to trap the particles, regardless of their
initial $y$-position or velocities, for
\begin{equation}\label{eq:equiBf}
  B_0 > \frac{1}{2} B_f.
\end{equation}
We now replace $B_f$ in the equation above by its expression from Eq.~(\ref{eq:equiBf}). We then express $\delta_W$ from Eq.~(\ref{eq:grW}) and find that $B_0 >   B_f/2$ is equivalent to,
\begin{equation}\label{eq:sigmaOK}
\sigma > \frac{\eta^2}{1 + \eta^2/2} ~ \beta_0^2 \equiv \kappa ~ \beta_0^2.
\end{equation}

The vertical asymptote defined by the blue curve on Fig.~\ref{fig:sigmagamma} relies therefore on the parameter $\eta$. The constraints on its value stem from the degree of equipartition reached at saturation of the Weibel instability. In this respect, let us compute the total amount of  magnetic energy contained in the magnetic field at saturation, and compare it to the upstream kinetic energy. Using Eq.~(\ref{eq:equi}) to express $B_f^2$ we find,
\begin{equation}\label{eq:equiTot}
\frac{(B_f^2/2 + B_0^2)/8\pi}{\gamma_0 n_0 m c^2} = \eta^2\beta_0^2+\frac{\sigma}{2} (1-\eta ^2).
\end{equation}
Accounting only for the energy contained in the Weibel field, one finds,
\begin{equation}\label{eq:equiW}
\frac{(B_f^2/2)/8\pi}{\gamma_0 n_0 m c^2} = \eta^2(\beta_0^2 - \sigma/2).
\end{equation}
In the relativistic regime where $\beta_0 \sim 1$, Eq.~(\ref{eq:equiTot}) gives unity for $\eta =1$, and is monotonically increasing for $\eta\in [0,1]$ and $\sigma < 2$. Note that the regime $\sigma > 2\beta_0^2$ is irrelevant in the present context since Eq.~(\ref{eq:grW}) implies the instability vanishes there. Eq.~(\ref{eq:equiW}) implies that the relative amount of energy contained in the Weibel field varies like $\eta^2$. For $\eta = 1$, Eq.~(\ref{eq:sigmaOK}) give $\kappa=2/3$, which is the values considered on Fig.~\ref{fig:setup}.


\end{document}